\documentstyle[openbib,12pt,fleqn]{article}
\newcommand{\bea}{\begin{eqnarray}}
\newcommand{\beq}{\begin{equation}}
\newcommand{\eea}{\end{eqnarray}}
\newcommand{\eeq}{\end{equation}}

\renewcommand{\baselinestretch}{2}

\begin{document}
\bibliographystyle{unsrt}

\setcounter{footnote}{0}

%
%
%
%
\begin{center}
\phantom{.}
{\Large \bf Particle propagation\\ in a random\&quasiperiodic potential \\}
{\small \sl   F.~BORGONOVI~$^{[a]}$\\}
{\small \it Dipartimento di Matematica, Universit\`a Cattolica,
via Trieste 17, 25121 Brescia, ITALY}

{\small  \sl D.L.~SHEPELYANSKY~$^{[b]}$ 
\\}
{\small \it Laboratoire de Physique Quantique, UMR C5626 du CNRS,\\
  Universit\'e Paul Sabatier, 31062 Toulouse Cedex, FRANCE}\\

\vspace{0.5truecm}

\vskip .3 truecm

\vspace{0.5truecm}
\end{center}
\small
{\bf Abstract:\/}
We numerically investigate the Anderson transition 
in an effective dimension $d$
($3 \leq d \leq 11$) for one particle propagation
in a model random\&quasiperiodic potential. The found critical exponents
are different from the standard scaling picture. 
We discuss possible reasons for this difference.
\vskip .6truecm

{PACS numbers {71.55.Jv, 05.45.+b }}
\newpage

The Anderson transition  has been intensively investigated during  last 
years. According to the scaling theory  for spinless particles 
all states are localized 
for dimension $d \le 2$, while a transition from localization
to diffusive propagation occurs for $d > 2$ 
when a certain hopping parameter $k$ crosses a critical value $k_{cr}$
(see reviews \cite{LR,MK}).
The one parameter scaling theory predicts
that in the   localized phase  ($ k < k_{cr}$) the
typical localization length  of wavefunctions  
 diverges at the critical point as
$l \sim \vert k - k_{cr} \vert^{-\nu}$.
Above the critical point the dynamics is diffusive 
and the $DC$ conductivity $\sigma$ 
which is proportional to the diffusion rate $D$ is assumed to 
approach zero at the critical point as 
$\sigma \propto D \sim \vert k - k_{cr}\vert^s$.
Scaling arguments give the following  relation between the   exponents:
$ s = (d-2)\nu$. For dimension $d=2+\epsilon$ the $\epsilon$ expansion
theory predicts $s=\nu=1+O(\epsilon^4)$\cite{Hikami}.
In higher dimensions the problem was studied by many authors 
\cite{Efetov,Zinbauer,Fyodorov}. According to 
the results presented
there $\nu = 1/2$ for $d\to \infty$ while for the delocalized 
phase $s \approx d/2$ \cite{Fyodorov} or $\sigma$ decreases
exponentially near the transition\cite{Efetov}.

Numerical investigations of the exponents has been restricted 
to $d=3$ where it was found $s=\nu=1.5\pm 0.1$ (\cite{MK} and 
refs. therein)  in agreement with the scaling relation
between $\nu$ and $s$.
However, the applied numerical methods were quite heavy 
and did not allow to obtain a better accuracy in the 
determination of the exponents or to increase the dimension.
For $2 < d \leq 3$ and $d=4$ there are only recent results 
for $\nu$ \cite{4dPRL}
obtained by transfer matrix technique. While the results there seem
to be in agreement with the scaling theory the system
size was so small that the question if the thermodynamic limit
had been reached remains open.

An effective way to increase the number of dimensions
was proposed in  \cite{SheD} and applied for investigation
of transition in $d=3$ \cite{CGS} where it was also found 
$s\approx 1.25 $ and $\nu\approx 1.5$.
The method consists in the investigation of 
the well known model of quantum chaos namely the kicked 
rotator model with a frequency--modulated amplitude of kicks.
The time--dependent Hamiltonian of the model is given by :
 
\begin{equation}
\begin{array}{c}
H=H_0(\hat{n}) + V(\theta,t) \delta_1(t)
\end{array}
\label{kr}
\end{equation}
where $\delta_1(t)$ is the periodic delta function with period 1
between kicks,   $\hat{n} = -i\partial/\partial \theta$ and
$\theta$ is a periodic angle variable.
$H_0$ determines the spectrum of unperturbed energies $E_n$  
chosen randomly distributed in $(0,2\pi)$. The perturbation $V$
depends on time in a quasi periodic way:

\begin{equation}
\begin{array}{c}
  V(\theta,t) =
  -2\tan^{-1} (2k (\cos\theta+\sum_{j=1}^{d-1} \cos (\theta_j+\omega_j t)))
\end{array}
\label{per}
\end{equation}
with $d-1$ incommensurate frequencies $\omega_j$. Here $\theta_j$
are initial phases and the time is measured in number of kicks.
The Hamiltonian (\ref{kr}) can be re--written in the  
 extended phase space by letting 
 $\hat{n}_j = -i \partial/\partial\theta_j$.
After that the problem becomes periodic in time and the 
eigenvalues equation for the quasi-energy eigenfunctions 
can be mapped to the usual solid--state form \cite{Fish,CGS}:
\begin{equation}
\begin{array}{c}
 T_{\bf n} u_{\bf n} + k \sum_{\bf r}' u_{\bf n-r}   = 0
\end{array}
\label{solid}
\end{equation}
where the $\sum'$ indicates a sum over the nearest neighbors to ${\bf n}$
on a $d$--dimensional lattice and  
\begin{equation}
\begin{array}{c}
 T_{\bf n} = -\tan ( 1/2 (E_n + \sum_{j=1}^{d-1} \phi_j  +\lambda))
\end{array}
\label{loyd}
\end{equation}
Here $\phi_j = n_j \omega_j$ and $\lambda $ is the quasi-energy.
If $\phi_j$ are randomly distributed in $(0,2\pi)$ then Eq. (\ref{solid})
becomes equivalent to the Lloyd model at the center of the band ($E=0$). 
The parameter $\lambda$  determines only the phase shift and it is clear that 
the physical characteristics are independent on its value.
Since the mapping between (\ref{kr}) and (\ref{solid}) is exact,
it is possible to study the Anderson transition in $d$--dimensions
by investigating the dynamics of the 1--dimensional system (\ref{kr}).
This gives an effective gain  of order $N^{d-1}$ in  numerical computations
if $N$ is the system size. Finally, discussing the model,
we should mention that the presence of disorder in the expression
for $ T_{\bf n}$ is crucial. Indeed, according 
to the exact mathematical results
\cite{Pastur} in the case of pure quasiperiodic potential
when in (\ref{loyd}) $E_n = \omega n$ all states 
for typical irrational frequencies are exponentially localized
for any $d$. The physical meaning of this result is quite clear:
the classical dynamics in this case is integrable and
variation of unperturbed actions (levels ${\bf n}$) is restricted by
invariant curves. However, even if only in one direction
the dispersion becomes nonlinear then the classical dynamics can
become chaotic with diffusive spreading in all ${\bf n}$ directions.
In this paper we investigate how this diffusion is affected by
quantum effects.

Using the above approach we studied numerically  the Anderson transition
for integer $3 \leq d \leq 11$ in the model (\ref{kr}) - (\ref{per}).
The choice of frequencies was the following:
for $d=3$ we fixed $\omega_{1,2}/2\pi = 1/\lambda, 1/\lambda^2$
with $\lambda = 1.3247 ...$ being the real root of
the cubic equation $x^3-x - 1 = 0$ which gives the most irrational pair
\cite{CGS}; for $d=4$ we added $\omega_3/2\pi = 1/\sqrt{2}$
and for $d > 4$ we chose all frequencies randomly in the interval
$(0,2\pi)$. The size of the basis $N$ was between 1024 and 4096.
The total number of iterations (kicks) was usually
around $10^6$ but in some cases close to the
critical point the evolution was followed up to
$10^7$ kicks. We used from 10 to 100 realizations of disorder
to suppress statistical fluctuations.

A typical example of diffusive spreading over the lattice for 
$d=4$ is shown in Fig.1. Here $k > k_{cr}$ and the second moment of the
probability distribution grows linearly with time. 
At the same time the probability distribution over levels has
the gaussian shape (see Fig. 1b). This allows to determine the diffusion constant $D$.
Usually, we extract it from the probability distribution
since here the statistical fluctuations are lower than for the
value obtained from the second moment $D=n^2/t$. However, 
both methods give quite close values. The value of $D$ found in this
way is then averaged over different realizations of disorder.
For $k < k_{cr}$ the probability distribution averaged in time 
reachs a stationary exponentially localized form and at the same
time the growth of the second moment $n^2$ is saturated (Fig.2).
From the obtained stationary distribution the localization length 
is determined in the two ways. One is by the square list fitting
of $\ln \vert \psi_n \vert^2 = - 2 \gamma n +b$ with the
localization length being $l=1/\gamma$ and $b$ some constant
\cite{rem}. Another definition  is via the participation ratio so that
$\gamma_i = {\sum_{n}} \vert \psi_n \vert^4/{\sum_{n}} \vert \psi_n \vert^2$.
After that both values $\gamma$ and $\gamma_i$
were averaged over different realizations of disorder.
The inverse participation ratio $1/\gamma_i$ determines another 
length scale which in principle can be parametrically different from $l$.

The numerical results for the dependence of $D, \gamma$ and $\gamma_i$
on parameter $k$ for the effective dimension $d=3$ are presented in Figs. 3,4.
To determine the scaling near the critical point we used
the 3 parameter fit of the type 
$\gamma_{(i)}=\gamma_0 \vert k - k_{cr} \vert^{\nu}$ 
and $D=D_0 \vert k - k_{cr} \vert ^s$.
The results of the fitting are given in the figures
captures (see also Table 1) as well as the parameters of 
$\chi^2$ test. Formally the statistical error of the exponents $s, \nu$
found in this way is rather small (less than 1\% of the value).
However, the estimate of non-statistical errors is quite difficult
since the fitting procedure is rather sensitive to the
value of $k_{cr}$. From comparison of $k_{cr}$ values obtained from
diffusive and localized phases it can be estimated on the level of 
5\%. The values of the exponents for $D$ and $\gamma_i$ are in 
the good agreement with the results of \cite{CGS} (see also \cite{rem}).
Our data indicate significant difference for the exponents
$\nu$ defined via $\gamma$ and $\gamma_i$ for $d=3$. 
To demonstrate the dependence near the critical point we fixed the
value of $k_{cr}$ defined from Fig. 3 and show the behaviour
in log-log scale near $k_{cr}$ in Fig.4. 
The two parameter fit with fixed $k_{cr}$ shown in Fig.4
gives similar values of the exponents $s, \nu$ as in the case of Fig.3.
The linear dependence of $\ln \gamma, \ln D$ on 
$\ln \vert 1 - k/k_{cr} \vert$ describes quite well the variation of
localization length and diffusion in one/two orders of magnitude.

The case with the dimension $d=4$ was investigated in a similar way.
The results are presented in Figs. 5,6. They definitely show 
stronger deviation from the scaling relation between 
exponents $\nu, s$. Especially pronounced is the small value
of $s$ which remains less than 2.

Inspite of this deviation from the scaling relation the
behaviour at the critical point is close to the standard expectations
\cite{LR,MK,Hikami,Efetov,Zinbauer,Fyodorov}. Indeed, at $k_{cr}$
the conductance has a finite critical value $g^*$. From another side
$g \sim E_c/\Delta$ where $E_c \sim D/L^2$ is the Thouless energy, 
$D$ is the diffusion coefficient
and $\Delta \sim B/L^d$ is the level spacing in a block of size L
with $B \sim 1$ being the band width.
Therefore, at $k_{cr}$ one has $D \sim B g^*/R^{d-2}$ where
$R \sim L$ is a typical length scale. At the same time
$D=R^2/t$ so that finally $R^d \sim B g^* t$. Our numerical data
for the values of $k$ close to $k_{cr}$ (Figs.7,8) show that 
this relation works. Formal fits give $R^d \sim t^\alpha$
with $\alpha =1.13 (d=3), 1.12 (d=4)$ close to the expected value.
We also analyzed the decay of the
average probability to stay at the origin $n_0$:
$P_0 (T) = {1\over T} \int_0^T dt \vert \psi_{n_0} (t) \vert^2$.
As can be seen in Figs 7b,8b it is characterized by a power law decay.
The numerically obtained
values of the power are not far from $1/d$ 
(Figs. 7,8). This indicates
that multifractal exponents are relatively small. This fact
is also confirmed by the rescaling of the probability distribution
$f_n=\vert \psi_n \vert^2$ at different moments of time (Fig.9). 
It shows that after rescaling $t^{1/d} f_n$ and $n/t^{1/d}$
the distribution remains approximately stationary in time, being however not
exactly exponential.

We also investigated the delocalization transition for
$d=5, 11$ (Figs. 10, 11). From the localized side the
transition was very sharp (very large $\gamma_0$) and 
it is not clear how accurate
are the critical exponents obtained in this region even if
the formal statistical error is quite small. 
Namely, the fits for $\gamma$ in the localized phase give $\gamma_0 = 257(28),
k_{cr}=0.214(1), \nu=2.32(4)$ with $\chi^2 = 0.18$ at $d=5$ and
$\gamma_0 = 2115(181), k_{cr}=0.107(2), \nu=2.55(2)$ 
with $\chi^2=0.04$ at $d=11$.
However, from the diffusive side the transition 
is going in a rather smooth way with the exponent
$s $ close to 2 being quite different from the 
expectations of scaling theory (see Figs. 10,11). The values of the exponents
for different dimensions are presented in the Table 1.
It definitely shows that the scaling relation $ s = (d-2) \nu$
does not work at all. Contrary to that our numerical data
indicate that for $d \gg 1$ the exponent $s$ approaches to its
limiting value $s \approx 2$. 
Finally, let us mention that the dependence of $k_{cr}$ on $d$
is quite close to $k_{cr} \approx 1/d$.
This type of behaviour can be expected since in the kick
potential all frequencies are mixed only if $k \sum_{j=1}^d \cos(\omega_j t)
\sim k_{cr} d \sim 1$. 

In conclusion, we studied the Anderson transition in a model
random \& quasiperiodic potential with effective dimension $d \geq 3$.
This model demonstrates quite many features which are the same as for 
the standard Anderson transition in a disordered $d$-dimensional
potential. For $d=2$ all states are localized and the localization
length grows exponentially with the decrease of disorder \cite{SheD}.
For $d=3$ the model has a transition from localization
to diffusion with the critical exponents close to expected \cite{CGS}.
However, for higher dimensions the exponents strongly
deviate from the expected scaling relation $s=(d-2)\nu$.
Contrary to that our numerical data show that
for $d \gg 1$ one has $s \approx 2$.
It is possible to give the following argument supporting $s =2$.
For $d \gg 1$ the critical value of the coupling goes to zero 
$k_{cr} \sim 1/d$. Therefore, the change of action is governed by
the equation
\begin{equation}
\begin{array}{c}
{\partial n/\partial t} \approx k \sin \theta \sum_{j=1}^d \cos{\omega_j t}
\end{array}
\label{dif}
\end{equation}
For $d$ going to infinity this sum gives the real diffusive
process in which $n^2 \sim k^2 t \sim D t$. Due to that
it is in some sense natural to expect that asymptotically for $d \gg 1$
the exponent $s$ approaches to 2. However, further investigations are
required to conclude whether the behaviour of the scaling exponents is or isn't
a peculiarity of the model under investigation. 
In our opinion, for the classical model chaos and diffusion in all
directions can appear even if the motion is nonlinear only in
one direction. Due to that we think that the quantum dynamics of the above
model should be quite similar to a real disordered system in dimension $d$.

We thank V.Kravtsov for useful discussions.
One of us (F.B.) is very grateful to the Laboratoire de Physique 
Quantique, UMR C5626 du CNRS, Univ. P.Sabatier, for the hospitality extended
to him during his visit when this work has been started and another (D.L.S.)
thanks the University at Como for hospitality during 
different stages of this work.

\vfill\eject
\renewcommand{\baselinestretch} {2}

\vfill\eject

\begin{table}
\caption{
}
\begin{tabular}{llllll}
d&s&$\nu(\gamma_i)$&(d-2)$\nu(\gamma_i)-s$&$\nu(\gamma)$&(d-2)$\nu(\gamma)-s$\\

3 & 1.25(1)  & 1.71(6) & 0.46 & 2.37(1) & 1.12  \\
4 & 1.52(5)  & 2.59(2) & 3.66 & 2.53(1) & 3.54  \\
5 & 2.04(3)  &  -      &   -   & 2.32   & 4.92  \\
11& 1.87(1)  &  -      &   -   & 2.55   & 21.0  \\
\end{tabular}
\label{table1}
\end{table}

\begin{figure}
\caption{(a) Behaviour of the second moment on time for $k=0.33$,
$d=4$ (diffusive side) .
 Dashed line is the fitted diffusion coefficient $D=0.0205$.
 Fourier basis is $N=1024$.
(b) Logarithm of the averaged distribution function 
between $t=0.995 \cdot 10^6$ and $t=10^6$ ($t$=number of kicks). 
Full line is the best fit gaussian which gives a diffusion coefficient
$D=0.0229$.
On x-axis we put $n^2$.}
\end{figure}

\begin{figure}
\caption{(a) Behaviour of the second moment on time for $k=0.24$,
$d=4$ (localized side). 
Basis is $N=1024$.
(b) Logarithm of the averaged distribution function 
between $t=0.3995 \cdot 10^7 $ and $t=0.4 \cdot 10^7 $. 
On x-axis we put $n$.
Line represents the fitted localized distribution with $l=15.4$.}
\end{figure}

\begin{figure}
\caption{Inverse localization length (left side) and diffusion rate
(right side) for $d=3$. Lines are the separate fits:
$ \gamma = \gamma_0 (k_{cr} - k )^\nu$ with 
$\gamma_0 = 8.01(1)$, $k_{cr} = 0.509(1)$ and $\nu = 2.37(1)$
with $\chi^2 = 10.5$
and 
$ D = D_0 ( k - k_{cr} )^s$ with 
$D_0 = 2.56(2)$, $k_{cr} = 0.479(1)$ and $s = 1.25(1)$
with $\chi^2 = 32.6$. 
A similar fit for $\gamma_i = \gamma_{0i} (k_{cr}-k)^{\nu_i}$
gives $\gamma_{0i} = 3.7(3)$, $k_{cr}=0.489(4)$, 
$\nu_i = 1.71(6)$ and $\chi^2 = 15.4$.
Here $10-100$ random configurations
have been iterated up to $10^7$ kicks. 
Data errors are within the symbol size.
}
\end{figure}

\begin{figure}
\caption{$d=3$. 
(a) Logarithm of diffusion rate vs 
$ ln(k/k_{cr}-1)$ where $k_{cr}$ is extracted from the separate fit
 (Fig.3). Straight line is the best fit line with  slope
$s = 1.248(5)$ and $\chi^2 = 10.6$.
(b) Logarithm of the inverse localization length vs 
$ ln(1-k/k_{cr})$ where $k_{cr}$ 
is extracted from the separate fit
(Fig.3). Straight line is the best fit line with  slope
$\nu = 2.374(9)$ and $\chi^2 = 33.8$.
(c)  Logarithm of the participation ratio $\gamma_i$ vs 
$ ln(1-k/k_{cr})$ where $k_{cr}=0.489(4)$ is extracted in 
a similar way from a three parameter fit.
Here the straight line has slope $\nu_i = 1.71(6)$ and
$\chi^2 = 16.3$
Here $10-100$ random configurations
have been iterated up to $10^7$ kicks. Errors are within the symbol size.}
\end{figure}

\begin{figure}
\caption{Inverse localization length (left side) and diffusion rate
(right side) for $d=4$. Lines are the separate fits:
$ \gamma = \gamma_0 ( k_{cr} - k )^\nu$ with
$\gamma_0 = 66.(10)$, $k_{cr} = 0.306(2)$ and $\nu = 2.53(1)$
with $\chi^2 = 10.2$
and
$ D = D_0 ( k - k_{cr} )^s$ with
$D_0 = 2.29(2)$, $k_{cr} = 0.283(3)$ and $s = 1.52(5)$
with $\chi^2 = 111$.
A similar fit for $\gamma_i = \gamma_{0i} (k_{cr}-k)^{\nu_i}$
gives $\gamma_{0i} = 85(28)$, $k_{cr}=0.305(5)$, 
$\nu_i = 2.59(2)$ and $\chi^2 =4.5$.
Here $10-100$ random configurations
have been iterated up to $10^7$ kicks.
Data errors are within the symbol size.
}
\end{figure}

\begin{figure}
\caption{$d=4$.
(a) Logarithm of diffusion rate vs
$ ln(k/k_{cr}-1)$ where $k_{cr}$ is extracted from the separate fit
(Fig.5). Straight line is the best fit line with  slope
$s = 1.519(4)$ and $\chi^2 = 115$.
(b) Logarithm of the inverse localization length vs
$ ln(1-k/k_{cr})$ where $k_{cr}$
is extracted from the separate fit
(Fig 5). Straight line is the best fit line with  slope
$\nu = 2.534(5)$ and $\chi^2 = 91.81$.
(c)  Logarithm of the participation ratio $\gamma_i$ vs
$ ln(1-k/k_{cr})$ where $k_{cr}=0.305(5)$ is extracted in
a similar way from a three parameter fit.
Here the straight line has slope $\nu_i = 2.59(2)$ and
$\chi^2 = 4.5$
Here $10-100$ random configurations
have been iterated up to $10^7$ kicks. Errors are within the symbol size.}
\end{figure}

\begin{figure}
\caption{$d=3$
Study near the critical point at k=0.48. (a) Behaviour 
of $R^d = \sum_{n} \vert n \vert^d \vert \psi_n \vert^2$ in time. 
Best fit (dashed line) gives   $ R^d \sim t^{1.13(3)} $.
(b) Behaviour of the logarithm of the 
integrated probability to stay $P_0(T)$
as a function
of the logarithm of time $T$. Here is 
$P_0 (T) = {1\over T} \int_0^T dt \vert \psi_{n_0} (t) \vert^2$ and 
the initial state vector is $\psi(t=0) = \delta_{n,n_0}$. 
Best  fit (dashed line) 
gives $P_0(T) \sim T^{-0.28(3)}$. One random configuration 
has been considered. Basis is N=4096.}
\end{figure}

\begin{figure}
\caption{The same as Fig.7 but for d=4 near the critical point 
at k=0.27.
Results from fit gives $R^d \sim  t^{1.12(3)} $ and $P_0(T) \sim T^{-.26(3)}$}
\end{figure}

\begin{figure}
\caption{Rescaling of the distribution function in the critical
region ($k=0.27$) 
for $d=4$. (a) The averaged (over $10^4$ kicks ) probability
distribution over the unperturbed basis $f_n$ taken at three different times 
$t_1, 2\ t_1, 3\ t_1$ where $t_1= 10^6$, shows that as the time goes on 
the distribution increases its size. 
(b) The same as (a) but in the rescaled variables
$t^{1/d} f_n$ and $n/t^{1/d}$ (since at the critical point
$\vert n \vert^d \sim t$
with $d=4$).}
\end{figure}

\begin{figure}
\caption{$d=5$. (a) Diffusion rate vs $k$,
3 parameters fit (full line) gives :
$D= 2.56(6) \ (k-.180(2))^{2.04(3)}$ with $\chi^2 = 36.1 $;
(b) logarithm of diffusion rate vs
$ ln(k/k_{cr}-1)$ where $k_{cr}$ is extracted from (a).
Best fit (full line) has the slope $s=2.039(9)$}
\end{figure}

\begin{figure}
\caption{$d=11$. (a) Diffusion rate vs $k$,
3 parameters fit gives :
$D= 1.35(1) \ (k-.0924(5))^{1.87(1)}$ with $\chi^2 = 310 $;
(b) logarithm of diffusion rate vs
$ ln(k/k_{cr}-1)$ where $k_{cr}$ is extracted from (a)
Best fit (full line) has the slope $s=1.99(5)$}
\end{figure}

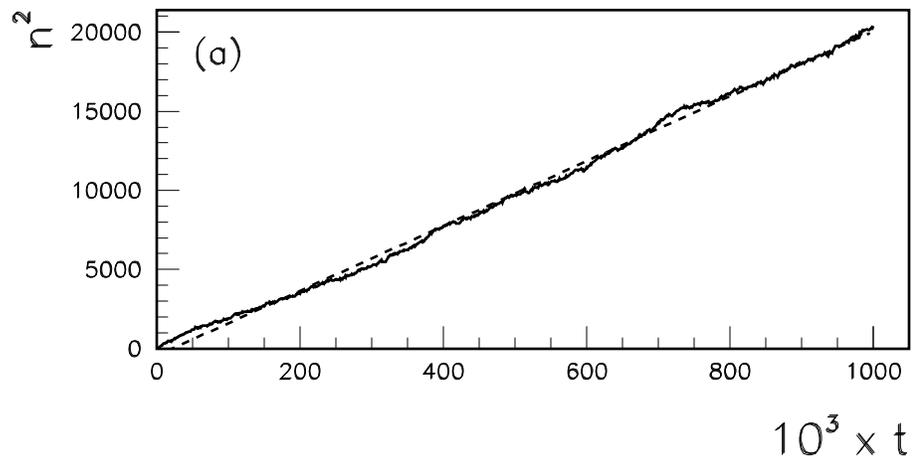
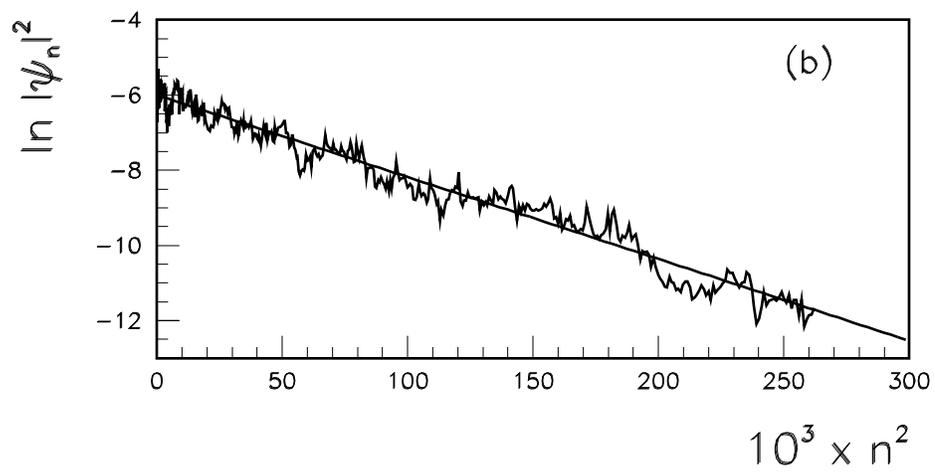
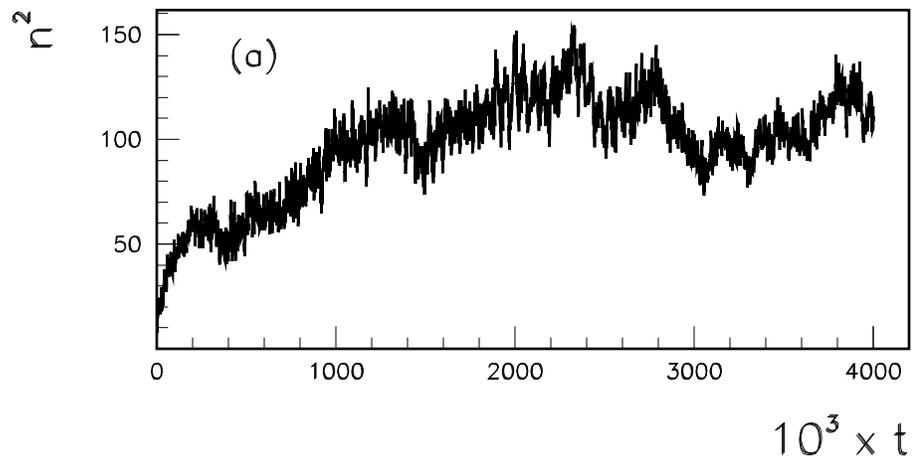
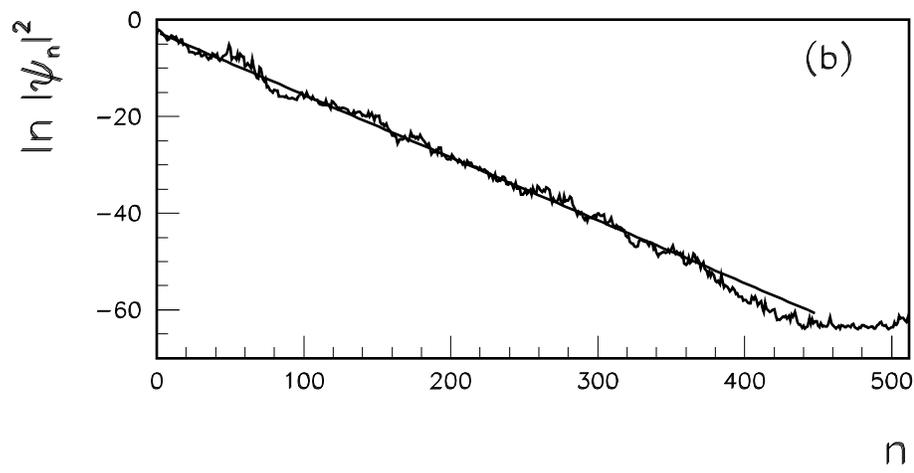
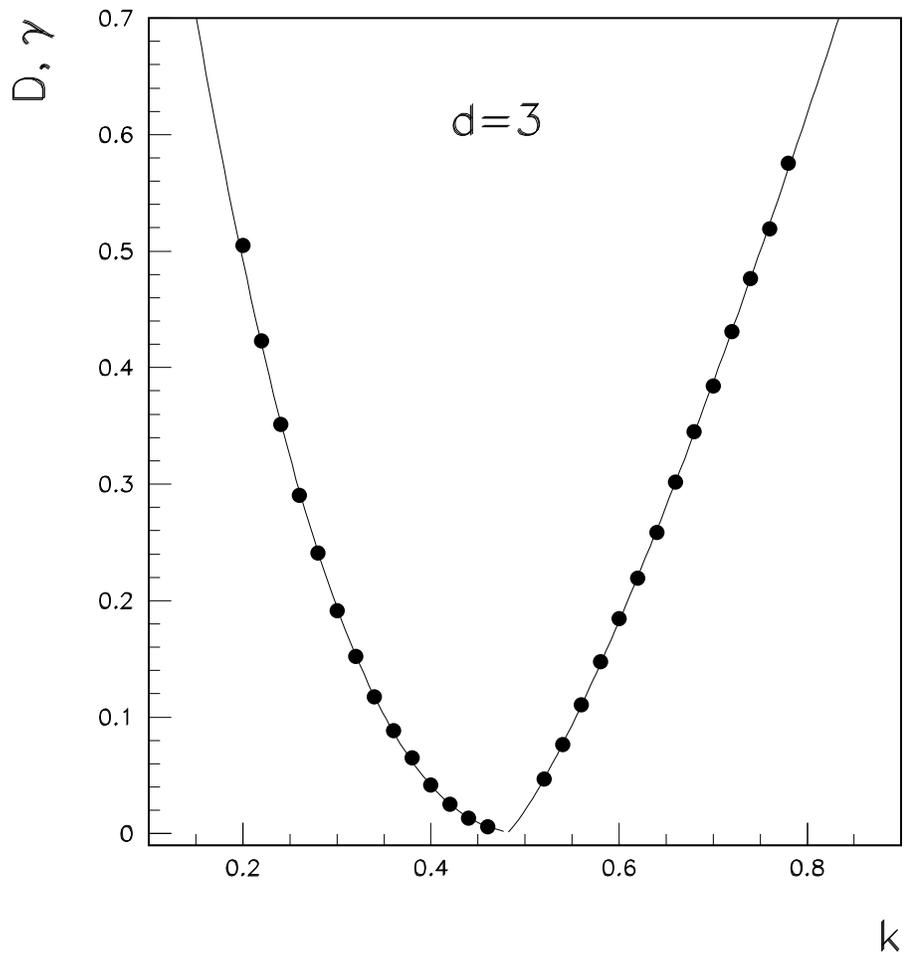
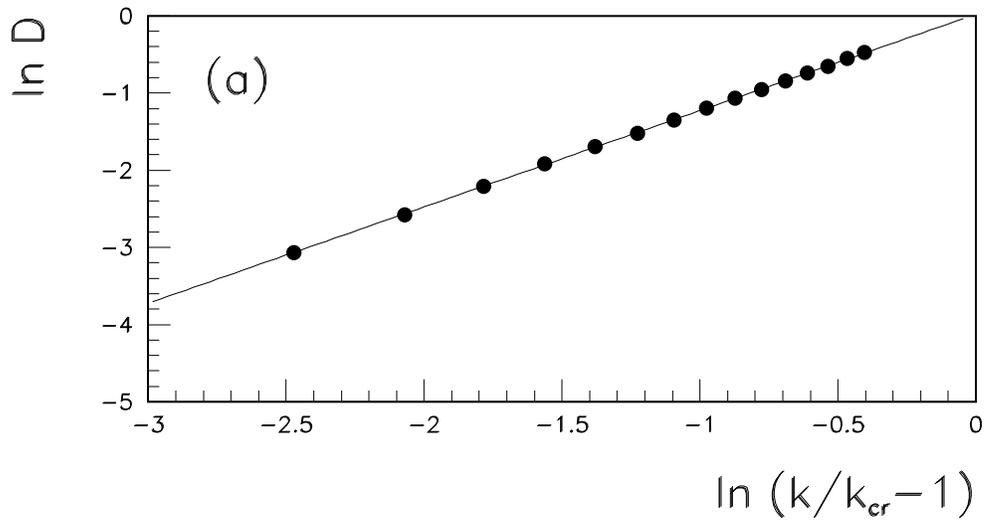
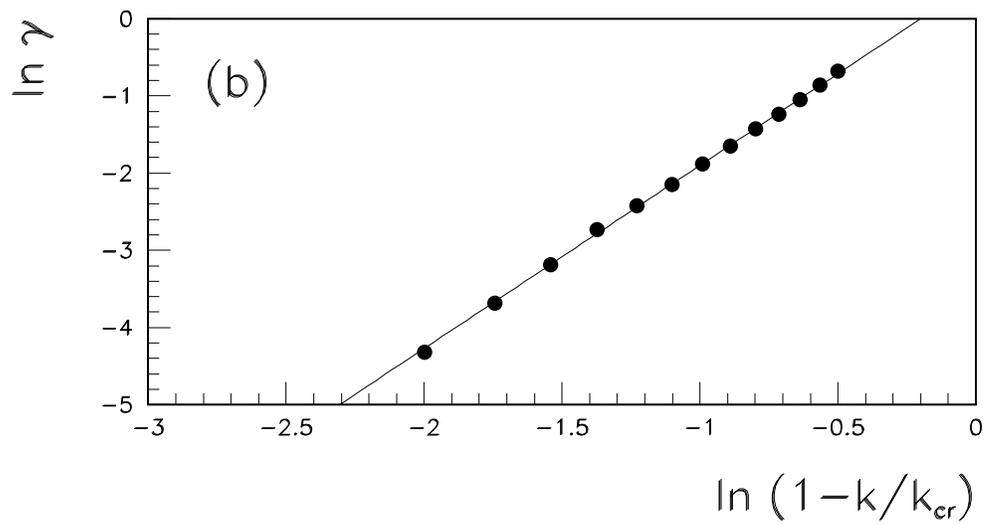
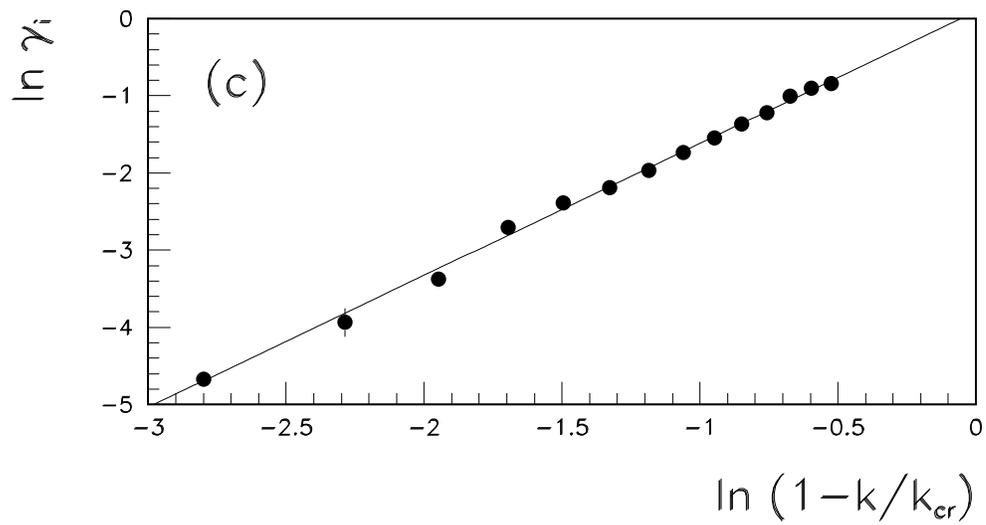
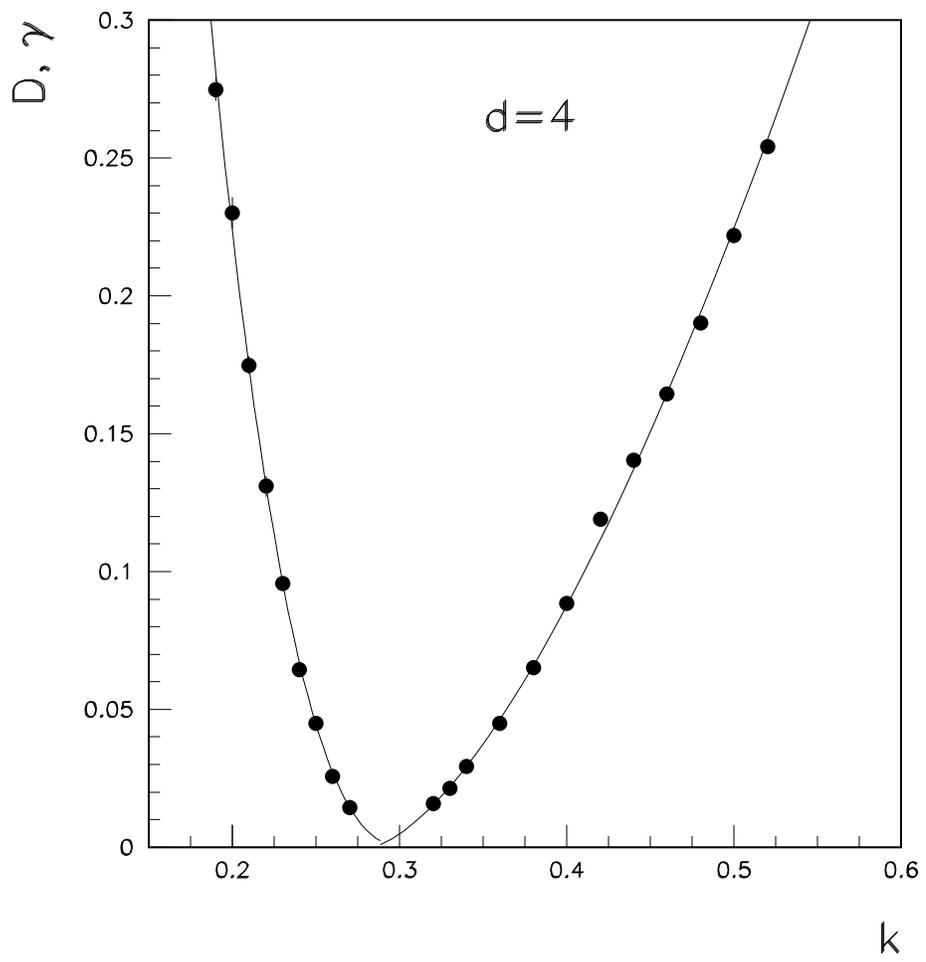
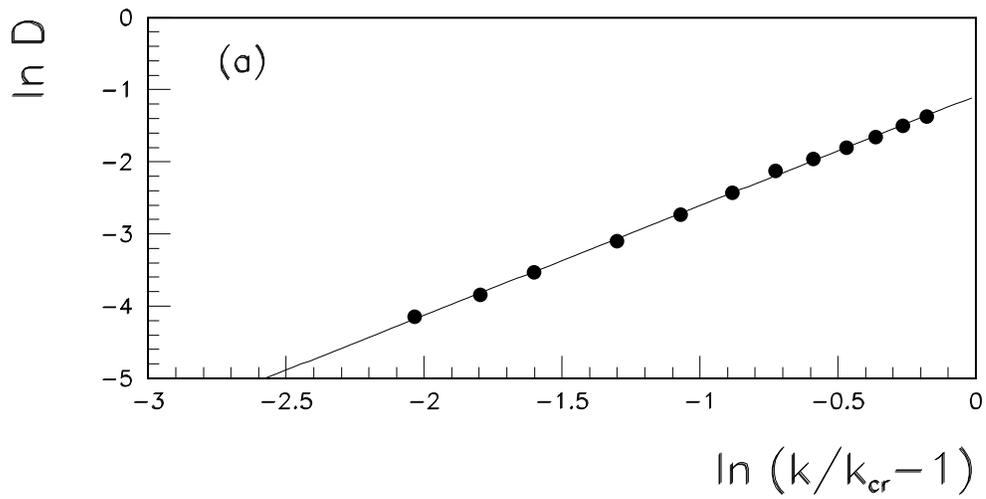
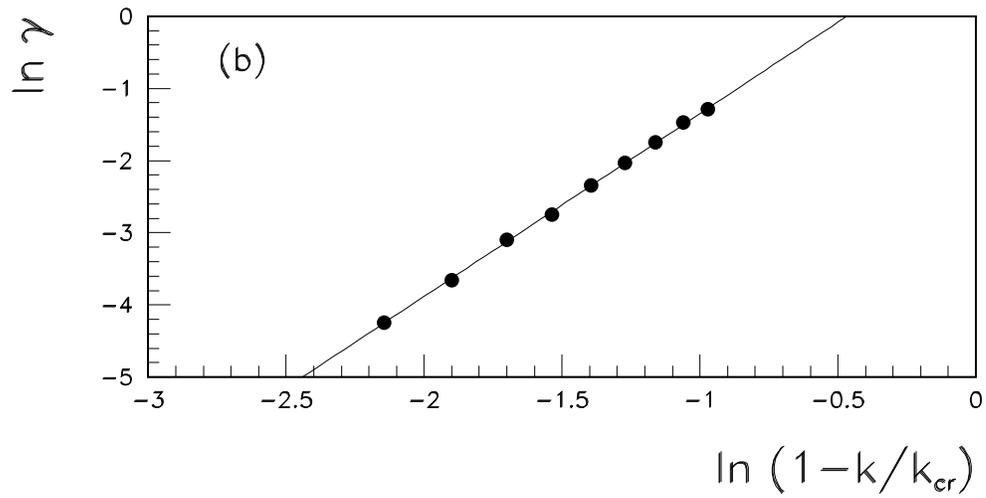
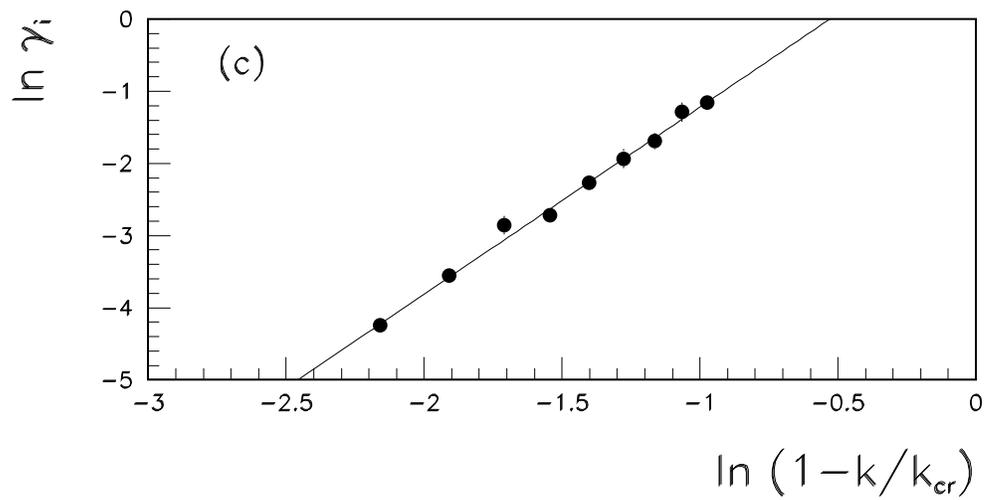
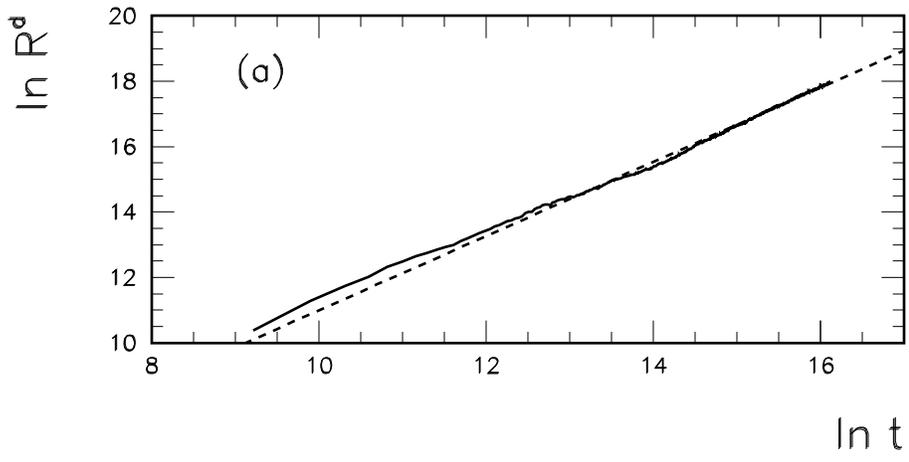
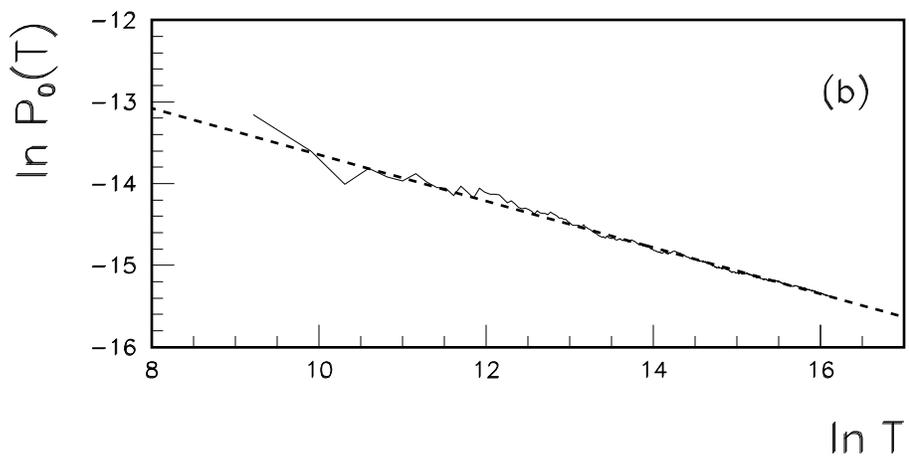
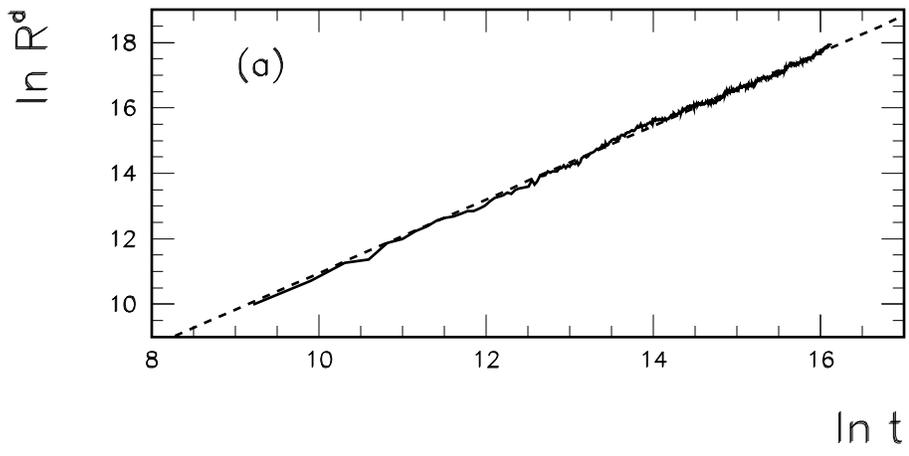
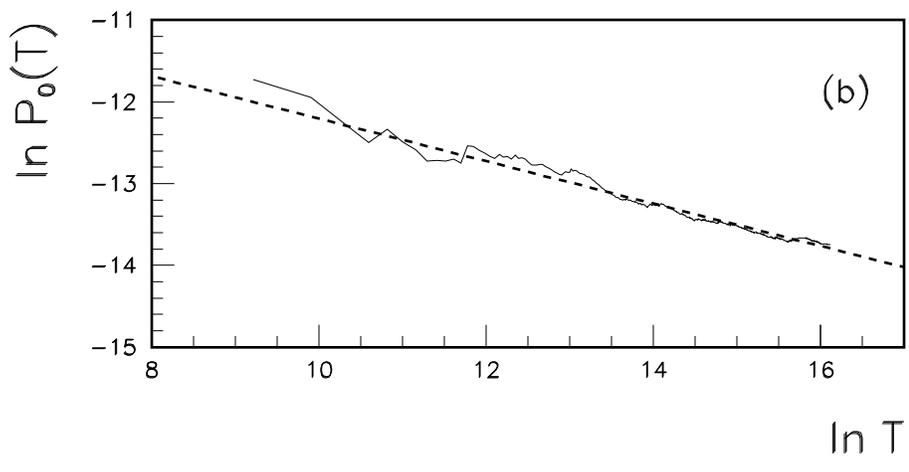
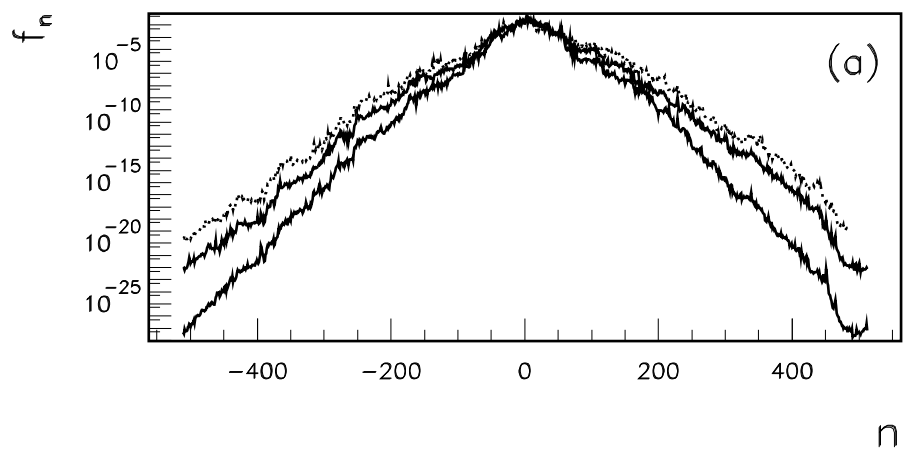
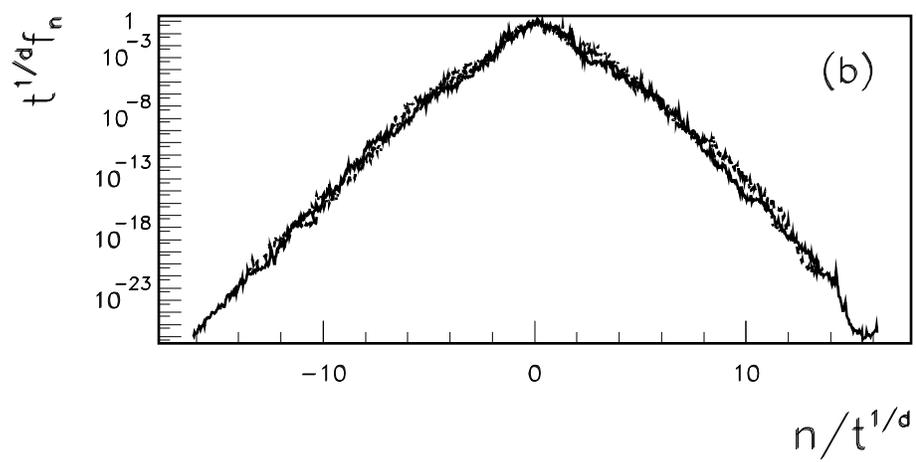
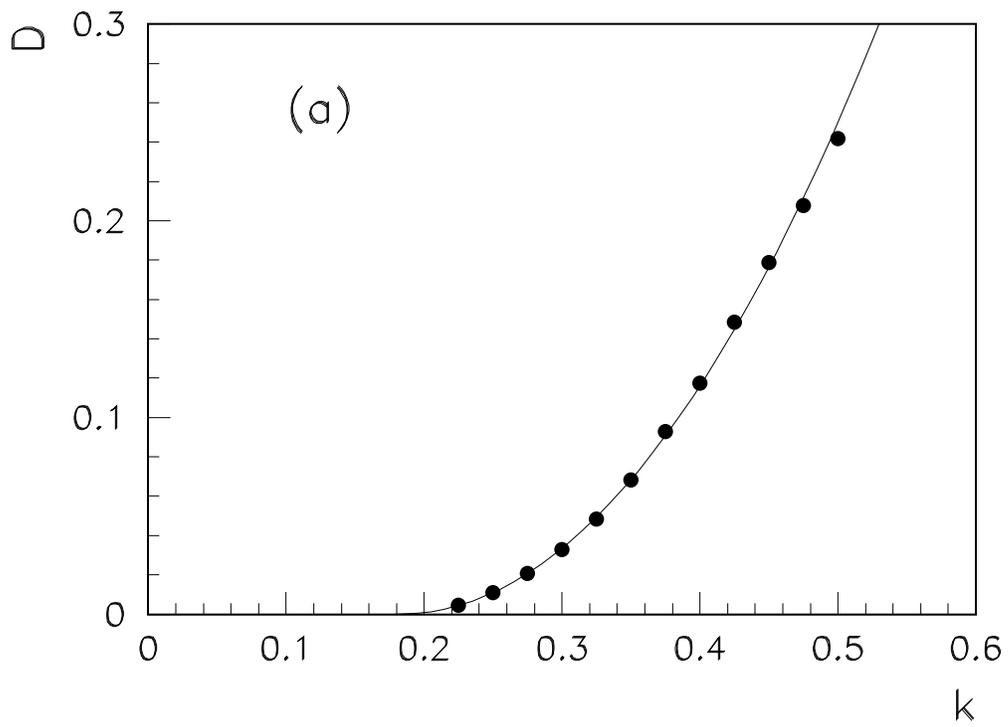
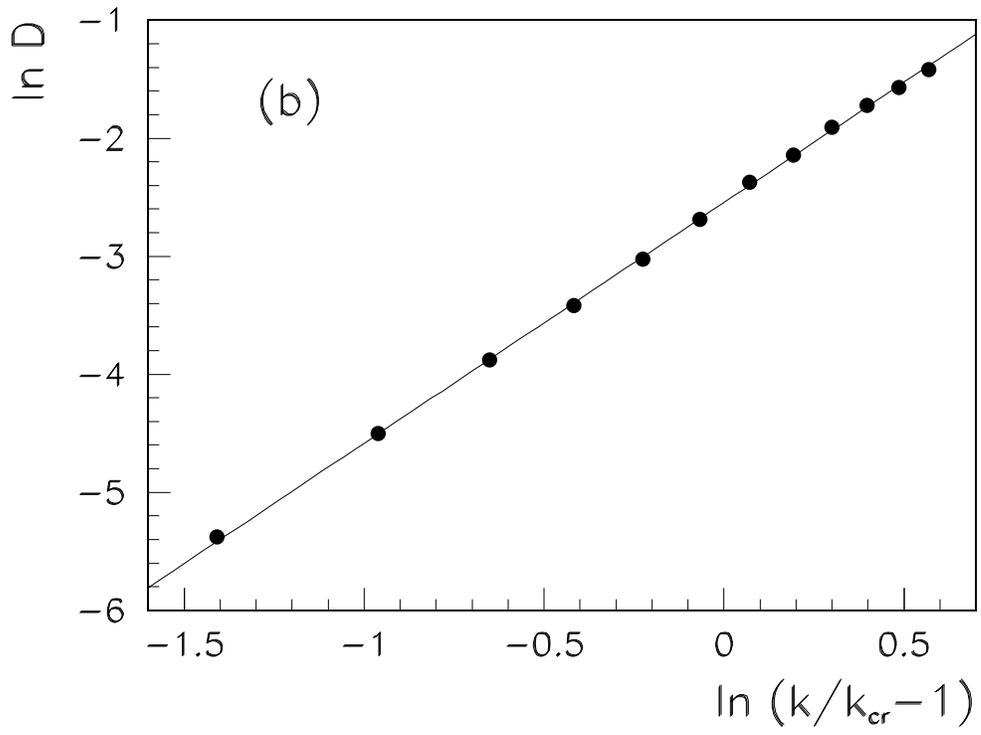
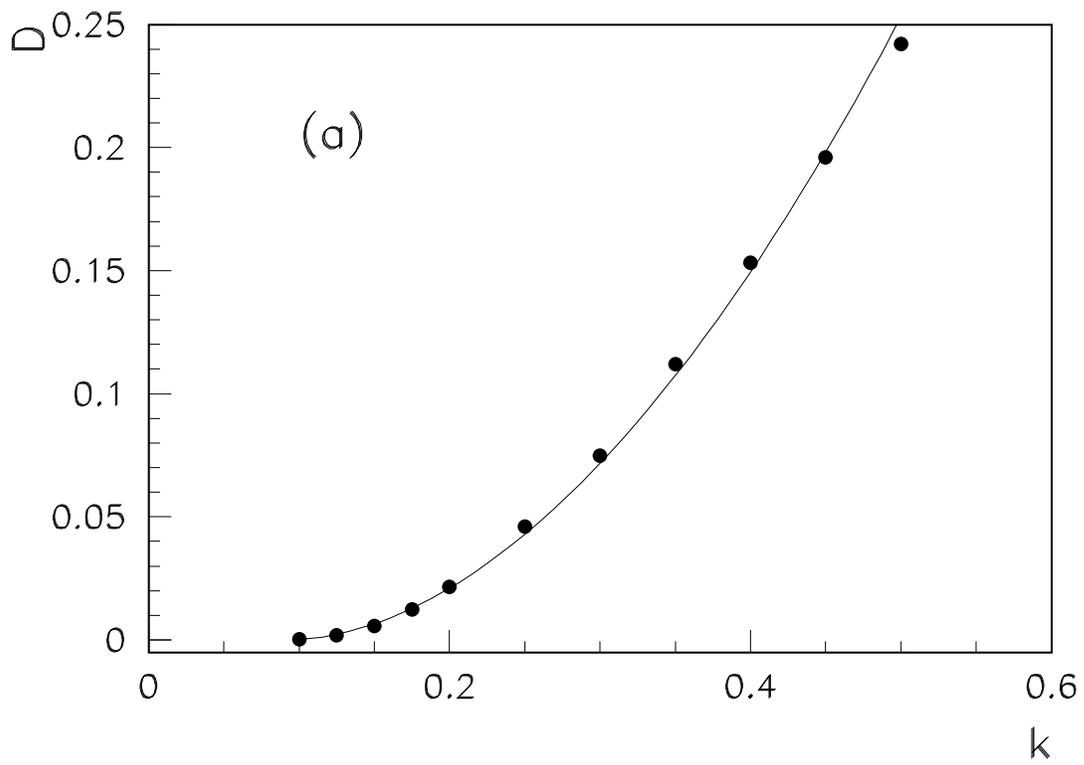
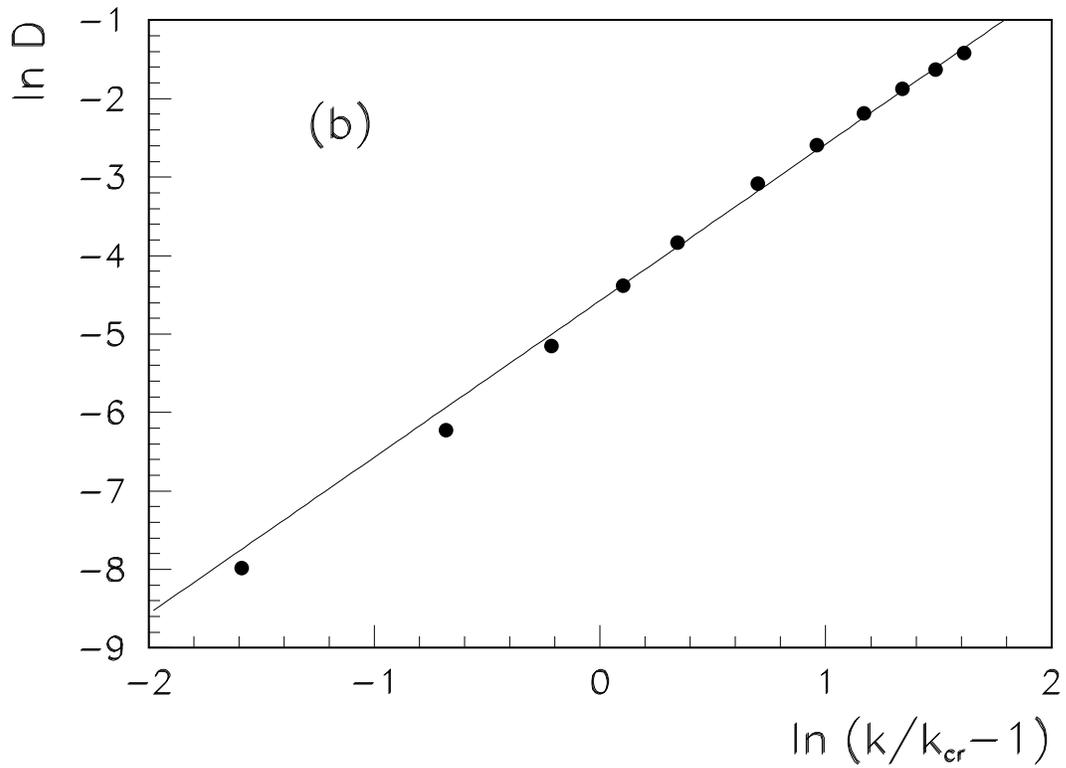
\end{document}